\documentclass[fleqn,10pt]{wlscirep}
\usepackage[utf8]{inputenc}
\usepackage[T1]{fontenc}
\usepackage{graphicx}
\usepackage{amssymb}
\usepackage{amsmath}
\usepackage{subfig}
\usepackage{color}
\usepackage{bm}

\title{An approach using geometric diagrams to generic Bell inequalities with multiple observables}
\author[1,2]{Junghee Ryu}
\author[3,4]{Jinhyoung Lee}
\author[5,*]{Hoon Ryu}
\affil[1]{Center for Quantum Information R\&D, Korea Institute of Science and Technology Information, Daejeon, 34141, Korea}
\affil[2]{Division of Quantum Information, KISTI School, Korea University of Science and Technology, Daejeon 34141, Republic of Korea}
\affil[3]{Department of Physics, Hanyang University, Seoul, 04763, Republic of Korea}
\affil[4]{Center for Quantum Simulation, Korea Institute of Science and Technology (KIST), Seoul, 02792, Republic of Korea}
\affil[5]{School of Computer Engineering, Kumoh National Institute of Technology, Gumi, Gyeongsangbuk-do, 39177, Republic of Korea}
\affil[*]{Corresponding author: elec1020@kumoh.ac.kr}

\begin{abstract}
We extend the generic Bell inequalities suggested by Son, Lee, and Kim [Phys. Rev. Lett. {\bf 96}, 060406 (2006)] to incorporate multiple observables for tripartite systems and introduce a geometric methodology for calculating classical upper bounds of the inequalities. Our method transforms the problem of finding the classical upper bounds into identifying constraints in linear congruence relations. Using this approach, we derive the upper bounds for scenarios with three and four observables per party. In order to demonstrate quantum violations, we employ Greenberger-Horne-Zeilinger entangled states that can achieve values exceeding the classical upper bounds, with the violation becoming more pronounced as the number of observables increases.
\end{abstract}
\begin{document}

\flushbottom
\maketitle

\newcommand{\bra}[1]{\left\langle#1\right\vert}
\newcommand{\ket}[1]{\left\vert#1\right\rangle}
\newcommand{\abs}[1]{\left\vert#1\right\vert}
\newcommand{\avg}[1]{\left\langle#1\right\rangle}
\newcommand{\pro}[2]{|\left<#1|#2\right>|^2}
\newcommand{\tmp}[2]{\left\vert\langle#1\right\vert#2\left\rangle\vert^2\right}
\newcommand{\hket}[1]{\left\vert#1\right)}
\newcommand{\textleft}[2]{{\vphantom{#2}}{_{#1}}#2}
\newtheorem{theorem}{Theorem}\newcommand{\circled}[1]{\text{\textcircled{\scriptsize #1}}}

\thispagestyle{empty}

Bell derived a constraint, in terms of statistical inequality, on correlations for two remote systems that local realistic theories must obey, and he showed that the constraint can be violated by quantum mechanics in case of two coupled spin-$1/2$ particles and suitable local measurements~\cite{bell1964einstein}. This is known as Bell's theorem, which states that local realistic theories cannot completely simulate quantum correlations. Later, Clauser, Horne, Shimony, and Holt (CHSH) improved Bell's original idea by proposing a statistical inequality that is more experimentally feasible~\cite{clauser1969proposed}. Following these theoretical developments, significant experimental efforts were undertaken to empirically verify the predictions of Bell's theorem across various physical platforms (for historical review, see Ref.~\cite{brunner2014bell}). Notably, in $2015$ and $2017$ loophole-free Bell tests were successfully implemented in different physical platforms~\cite{hensen2015loophole, giustina2015significant, shalm2015strong, rosenfeld2017event}. Studying Bell inequalities has thus become fundamental to illustrating the profound differences between quantum mechanics and classical physics. Nowadays, it is well established that violations of Bell inequalities are essential for enabling various quantum information protocols to outperform their classical counterparts, such as random number generation, quantum key distribution, reducing communication complexity, and so on~\cite{horodecki2009quantum}, driving extensive research across both theoretical and experimental domains~\cite{brunner2014bell}. Recently, violations of Bell inequalities become central to self-testing, where certifying a unique quantum entangled state and local observables for quantum violations, as well as the corresponding classical upper bounds, is essential for constructing self-testing in a device-independent manner~\cite{mayers2003self, kaniewski2016analytic, kaniewski2017self, vsupic2020self, panwar2023elegant}.

The simplest case of Bell's theorem involves two spin-$1/2$ particles, where each subsystem is measured with two observables, and each observable yields two possible outcomes. This scenario can be represented using the notation $(N, M, d)$, where $N$ denotes the number of subsystems, $M$ represents the number of observables per subsystem, and $d$ specifies the number of outcomes per observable. Under this notation, the cases considered by Bell and CHSH correspond to $(N, M, d) = (2, 2, 2)$. Since Bell's discovery, many efforts have been devoted to extending Bell inequalities to more general scenarios, including multi-particle systems and configurations with more measurement outcomes. These investigations have led to significant progress in generalizing Bell inequalities to diverse physical contexts. For the $(N, 2, 2)$ case, Mermin, Ardehali, Belinskii, and Klyshko derived a set of inequalities~\cite{mermin1990extreme, ardehali1992bell, belinskiui1993interference}. For the $(2, 2, d)$ case, Collins, Gisin, Linden, Massar, and Popescu proposed inequalities\cite{collins2002bell}, demonstrating that quantum violations increase with the dimensionality of the Hilbert space, consistent with numerical results~\cite{kaszlikowski2000violations}. Extending this to the $(N, 2, d)$ case, Son, Lee, and Kim (SLK) proposed a set of Bell inequalities, referred to as generic in their work~\cite{son2006generic}, meaning that these inequalities are derived from the all-versus-nothing contradiction introduced by Greenberger, Horne, and Zeilinger (GHZ)\cite{kafatos2013bell}. Beyond this, significant research efforts have been devoted to generalizing Bell inequalities to various other scenarios~\cite{dilley2018more, bancal2013definitions, chen2002maximal, bancal2011detecting, zhu2021less, svetlichny1987distinguishing}.

While the SLK inequalities addressed the case of $M=2$, they left open the question of extending the approach to arbitrary $M$. In this paper, we generalize the SLK inequalities to incorporate multiple observables for tripartite systems, focusing on the cases of $M=3$ and $M=4$, while also discussing arbitrary values of $M$. This extension is particularly significant because the original SLK inequalities with $M=2$ are unable to demonstrate the quantum violations in odd-dimensional systems. Although a variant form of the SLK inequality was proposed to overcome this issue, it no longer retains the generic property of the original construction. In contrast, our generalization with increased values of $M$ not only overcomes this limitation but also preserves the generic structure of the inequalities. As a result, it enables the detection of quantum violations in previously inaccessible odd-dimensional cases. To this end, we first propose a quantum operator involving multiple observables for calculating the quantum expectation values, and demonstrate the expectation values obtained by maximally entangled states. Then, to establish the quantum violations, we prove that no local realistic description can achieve these quantum values. In principle, the classical upper bounds can be calculated by analyzing the structure of the convex polytope under the local realistic description. However, this method becomes computationally intractable for multipartite and high-dimensional systems. In order to calculate the upper bounds, we introduce a geometric methodology. Our method transforms the problem of finding the upper bounds into identifying constraints in linear congruence relations. Although the quantum violations for $M=4$ were previously established in Ref.~\cite{ryu2008bell}, we rederive these results using our geometric method, obtaining consistent results and thereby demonstrating the versatility of our approach. Our generalized Bell inequality can also be considered a generic Bell inequality because its underlying derivation is based on the all-versus-nothing contradiction shown in Ref.~\cite{ryu2014multisetting}. The geometric approach is not intended for a direct comparison with polytope-based methods in terms of computational efficiency, but rather offers conceptual and structural insights into the local realistic bounds.

\section*{Generic Bell operator with multiple observables}
Suppose that three observers, mutually separated at a long distance, share a $d$-dimensional generalized GHZ state of three subsystems in the form of 
\begin{equation}
\ket{\psi}=\frac{1}{\sqrt{d}} \sum_{n=0}^{d-1} \ket{n,n,n},
\label{eq:3_ghz}
\end{equation}
where $\{\ket{n}\}$ is an orthonormal basis set. In $d$-dimensional Hilbert space, an orthogonal measurement is represented by an operator $\hat{A}=\sum_{n=0}^{d-1} \lambda_n \ket{n}\bra{n}$ with an eigenvalue $\lambda_n$ and an eigenvector $\ket{n}$.
For the observables of the higher dimensional systems, we shall use the unitary observable, which have been widely employed in Bell's theorem investigations for higher dimensional systems~\cite{collins2002bell, lee2006probabilistic, ryu2008bell, son2006generic, cerf2002greenberger}. To this end, the observable operator $\hat{A}$ is assumed to take, as its eigenvalue, one of the values in $\{1,\omega,\dots,\omega^{d-1} \}$, where the values represent the $d$th roots of unity over the complex field, i.e., $\omega = \exp(2 \pi i /d)$. Then, it turns out that the operator $\hat{A}$ is unitary~\cite{cerf2002greenberger}.


Let us briefly outline the construction of unitary observables employed in our Bell scenario, as introduced in Ref.~\cite{son2006generic}. The method relies on applying quantum Fourier transformation and phase shift operations to a reference unitary observable. For our reference observable, we adopt $\hat{Z}=\sum_{n=0}^{d-1} \omega^{n} \ket{n}\bra{n}$. The observable operator $\hat{X}$ is obtained by applying the quantum Fourier transformation as
\begin{equation}
\hat{X}=\hat{F}\hat{Z}\hat{F}^{\dagger}=\sum_{n=0}^{d-1}\ket{n+1}\bra{n},
\label{eq:obs_x}
\end{equation}
where $\ket{n} \equiv \ket{n \mod d}$ and the eigenvector of $\hat{X}$ is given by
\begin{equation}
\ket{n}_x = \hat{F} \ket{n} = \frac{1}{\sqrt{d}} \sum_{m=0}^{d-1} \omega^{-nm} \ket{m}.
\end{equation}
The operator $\hat{X}$ shifts a basis state periodically: $\ket{n} \rightarrow \ket{n+1}$ and $\ket{d-1} \rightarrow \ket{0}$. For the operator $\hat{Y}$, the phase shift operation $\hat{P}_{1/2}=\sum_n \omega^{-n/2} \ket{n} \bra{n}$ is applied to the $\hat{X}$, i.e., $\hat{P}_{1/2} \hat{X} \hat{P}^{\dagger}_{1/2}$, and consequently the observable $\hat{Y}$ reads
\begin{equation}
  \hat{Y} = \omega^{-1/2} \left( \sum_{n=0}^{d-2} \ket{n+1} \bra{n} - \ket{0} \bra{d-1} \right).
  \label{eq:obs_y}
\end{equation}
The eigenvector of $\hat{Y}$ is given by
\begin{equation}
\ket{n}_y = \hat{P}_{1/2} \hat{F}\ket{n} =\frac{1}{\sqrt{d}} \sum_{m=0}^{d-1} \omega^{-(n+1/2)m} \ket{m}.
\end{equation}
The operator $\hat{Y}$ shifts a basis state as $\ket{n} \rightarrow \omega^{-1/2} \ket{n+1}$ and $\ket{d-1} \rightarrow -\omega^{-1/2} \ket{0}$. The operators $\hat{X}$, $\hat{Y}$, and $\hat{Z}$ reduce to the Pauli operators $\hat{\sigma}_x$, $\hat{\sigma}_y$, and $\hat{\sigma}_z$, respectively, when $d=2$. To increase the number of local observables, we employ the phase shift operations with rational phase values $\nu$ as $\hat{P}_{\nu}=\sum_n \omega^{-\nu n} \ket{n} \bra{n}$. Applying the $\hat{P}_{\nu}$ to $\hat{X}$ reads
\begin{equation}
\hat{X}(\nu)=\hat{P}_{\nu} \hat{X} \hat{P}^{\dagger}_{\nu} = \omega^{-\nu} \left( \sum_{n=0}^{d-2} \ket{n+1} \bra{n} + \omega^{ \nu d} \ket{0} \bra{d-1} \right).
\end{equation}
The observable $\hat{X}(\nu)$ shifts the basis states $\ket{n}$ as
\begin{equation}
  \hat{X}(\nu) \ket{n} = \begin{cases}
  \omega^{-\nu(1-d)} \ket{0} & \text{for $n=d-1$}\\
  \omega^{-\nu} \ket{n+1} & \text{otherwise}.
  \end{cases}
\end{equation}
Moreover, the generic Bell operator is constructed with components including the $n$-level raising operator $\hat{X}^n(\nu)$, which shifts as
\begin{equation}
  \hat{X}^{ n}(\nu) \ket{m} = \begin{cases}
  \omega^{-\nu(n-d)} \ket{m+n} & \text{for $m \geq d-n$} \\
  \omega^{-\nu n} \ket{m+n} & \text{otherwise}.
  \end{cases}
  \label{eq:Mobs}
\end{equation}
To generate a set of $M$ distinct observables, we utilize phase values $\nu$ in the range $\{0, 1/M,\cdots,(M-1)/M\}$. In case of $M=2$, the operator $\hat{X}(\nu)$ for $\nu=0$ reduces the $\hat{X}$ in Eq.~(\ref{eq:obs_x}) and $\hat{X}(1/2)$ reduces the $\hat{Y}$ in Eq.~(\ref{eq:obs_y}).

We formulate a generic Bell operator composed of the observables $\hat{X}(\nu)$ for tripartite $d$-dimensional systems. For our analysis involving three measurement settings per party, we select the phase values from the set $\{0,1/3,2/3\}$. Utilizing these observables $\hat{X}(\nu)$ along with their corresponding powers $\hat{X}^{n}(\nu)$, we construct the Bell operator for a $(3,3,d)$ system as follows:
\begin{equation}
  \hat{B}_{3} = \frac{1}{3^3} \sum_{n=1}^{d-1} \sum_{\gamma=0}^{2} \bigotimes_{j=1}^{3} \sum_{\eta_{j}=0}^{2} \Omega^{\gamma \eta_{j}} \omega^{n \eta_{j} /3} \hat{X}_{j}^{n}(\eta_{j} /3),
  \label{eq:3_bell_operator}
\end{equation}
where $\Omega = \exp(2 \pi i/3)$ and $\omega = \exp(2 \pi i/d)$. Note that $n$th powers operator $\hat{X}_{j}^{n}(\eta_{j} /3)$ is the $\eta_j$th observable of $j$th party. The expectation is given by
\begin{eqnarray}
  \bra{\psi}\hat{B}_3\ket{\psi} &=& \frac{1}{3^3} \sum_{n=1}^{d-1} \sum_{\gamma=0}^{2} \sum_{ \vec{\eta}=0}^{2} \Omega^{\gamma \tilde{\eta}} \omega^{n \tilde {\eta}/3} E^{n}(\eta_{1},\eta_{2},\eta_{3}) \nonumber \\
  &=& \frac{1}{3^{2}} \sum_{n=1}^{d-1}  \sum_{ \vec {\eta}=0}^{2} \delta(\tilde{\eta} \equiv 0 \mod 3)\omega^{n \tilde {\eta}/3} E^{n}(\eta_{1},\eta_{2},\eta_{3}),
 \label{eq:3bellQM}
\end{eqnarray}
where $ \sum_{ \vec {\eta}=0 } \equiv \sum_{\eta_{1}=0} \sum_{\eta_{2}=0} \sum_{\eta_{3}=0}$, $ \tilde{\eta} \equiv \sum_{j} \eta_{j}$, and $E^{n}(\eta_{1},\eta_{2},\eta_{3})$ is the $n$th order correlation function in the form of
\begin{equation}
E^{n}(\eta_{1},\eta_{2},\eta_{3})= \bra{\psi} \hat{X}_{1}^{n}(\eta_{1} /3) \otimes \hat{X}_{2}^{n}(\eta_{2} /3) \otimes \hat{X}_{3}^{n}(\eta_{3} /3) \ket{\psi}.
\end{equation}
Since the operators $\hat{X}_{j}$ and their powers $\hat{X}_{j}^{n}$ are all unitary, we have $\abs{E^n} \leq 1$. Consequently, the upper bound of Eq.~(\ref{eq:3bellQM}) can be achievable with the generalized GHZ state $\ket{\psi}$ in Eq.~(\ref{eq:3_ghz}) as
\begin{equation}
  \bra{\psi}\hat{B}_{3} \ket{\psi} = d-1.
  \label{eq:quantum_upperbound}
\end{equation}
This result follows from the fact that the generalized GHZ state is an eigenstate of the composite observables $\hat{X}_{1}^{n}(\eta_{1} /3) \otimes \hat{X}_{2}^{n}(\eta_{2} /3) \otimes \hat{X}_{3}^{n}(\eta_{3} /3)$ in Eq.~(\ref{eq:3_bell_operator}), as further elaborated in Ref.~\cite{ryu2014multisetting}. In the Appendix B, we show that the generalized GHZ state is the eigenstate of the Generic Bell operator.

\section*{Results and discussion}
Local hidden variables (LHVs) theories assume that the measurement outcomes are predetermined prior to actual measurement, with physical influences propagating no faster than the speed of light. Within this framework, the LHVs theories characterize quantum correlations of composite measurements as
\begin{equation}
\int d \lambda \rho(\lambda) \prod_{j=1}^{3} X_j (\lambda),
\end{equation}
where $\rho(\lambda)$ is the statistical distribution of hidden variables $\lambda$ with $\rho(\lambda) \geq 0$ and $\int d \lambda \rho(\lambda)=1$. The outcome of measurement $X_j(\eta_j/3)$ is predetermined as its eigenvalue $\omega^{\alpha(j,\eta_j)}$, where $\alpha(j, \eta_{j})$ is integer. As a result, the Bell function based on LHVs is given by
\begin{equation}
B_{3}^{\text{LHV}} = \frac{d}{3^{2}} \sum_{ \vec {\eta}=0}^{2} \delta_{3}(\tilde{\eta}) \delta_{d}(\tilde{\eta} /3 + \tilde{\alpha}) -1 \equiv  \mathcal{L} d -1,
  \label{eq:classical_function}
\end{equation}
with $\tilde{\eta} = \sum^{3}_{j=1}\eta_{j}$ and $\tilde{\alpha} = \sum^{3}_{j=1}\alpha(j,\eta_{j})$.
Here, $\delta_d (\alpha)=1$ if $\alpha \equiv 0 \mod d$ and $\delta_d (\alpha)=0$ otherwise. The explicit derivation of Eq.~(\ref{eq:classical_function}) is presented in the Appendix A, and the function $\mathcal{L}$ can be rewritten as
\begin{eqnarray}
 \mathcal{L} &=& \frac{1}{3^2} [\delta_d (a_{1} + a_{2} + a_{3}) + \delta_d (a_{1} + b_{2} + c_{3} +1) + \delta_d (a_{1} + c_{2} + b_{3} +1) \nonumber \\
&+& \delta_d (b_{1} + a_{2} + c_{3} +1) + \delta_d (b_{1} + b_{2} + b_{3} +1) + \delta_d (b_{1} + c_{2} + a_{3} +1) \nonumber \\ 
 &+& \delta_d (c_{1} + a_{2} + b_{3} +1) + \delta_d (c_{1} + b_{2} + a_{3} +1) + \delta_d (c_{1} + c_{2} + c_{3} +2) ].
\label{eq:LR_3}
\end{eqnarray}
Here we employ alternative notations with $a_j$, $b_j$, and $c_j$, where $a,b$, and $c$ indicate the $1$st, $2$nd, and $3$rd observables, respectively, and the subscript $j$ denotes the $j$th party. In other words, these correspond to the following replacements: $\alpha(j, 0) \rightarrow a_j$, $\alpha(j, 1) \rightarrow b_j$, and $\alpha(j, 2) \rightarrow c_j$. The maximum of $B_{3}^{\text{LHV}}$ depends on $\mathcal{L}$, which satisfies the constraint $0 \leq \mathcal{L} \leq 1$. When $\mathcal{L} =1$, the classical upper bound becomes $B_{3}^{\text{LHV}} = d-1$, which is equivalent to the quantum upper bound, resulting in no quantum violation. Consequently, the quantum violations can be observed whenever $\mathcal{L} < 1$.

To characterize these violations systematically, we propose a geometrical approach to determine the classical upper bound. To demonstrate our method, we first apply it to the simpler $(3, 2, d)$ system~\cite{son2006generic}. In this case, the Bell function based on LHVs is given by
\begin{eqnarray}
B_{2}^{\text{LHV}} = \frac{d}{4} \left[\delta_d (a_1 + a_2 + a_3) + \delta_d (b_1 + a_2 + b_3+1) + \delta_d (b_1 + b_2 + a_3+1) + \delta_d (a_1 + b_2 + b_3+1)\right]-1, 
\label{eq:LR_2}
\end{eqnarray}
and it consists of four delta functions with variables $a_i$ and $b_j$. It is shown that no values of $a_i$ and $b_j$ can simultaneously satisfy all four delta functions, with three being the maximum number that can be satisfied~\cite{son2006generic}. Consequently, the classical upper bound equals $3d/4 -1$. We reproduce the result by using our geometrical approach.

\begin{figure}[t]
  \centering
  \includegraphics[width=0.7\linewidth]{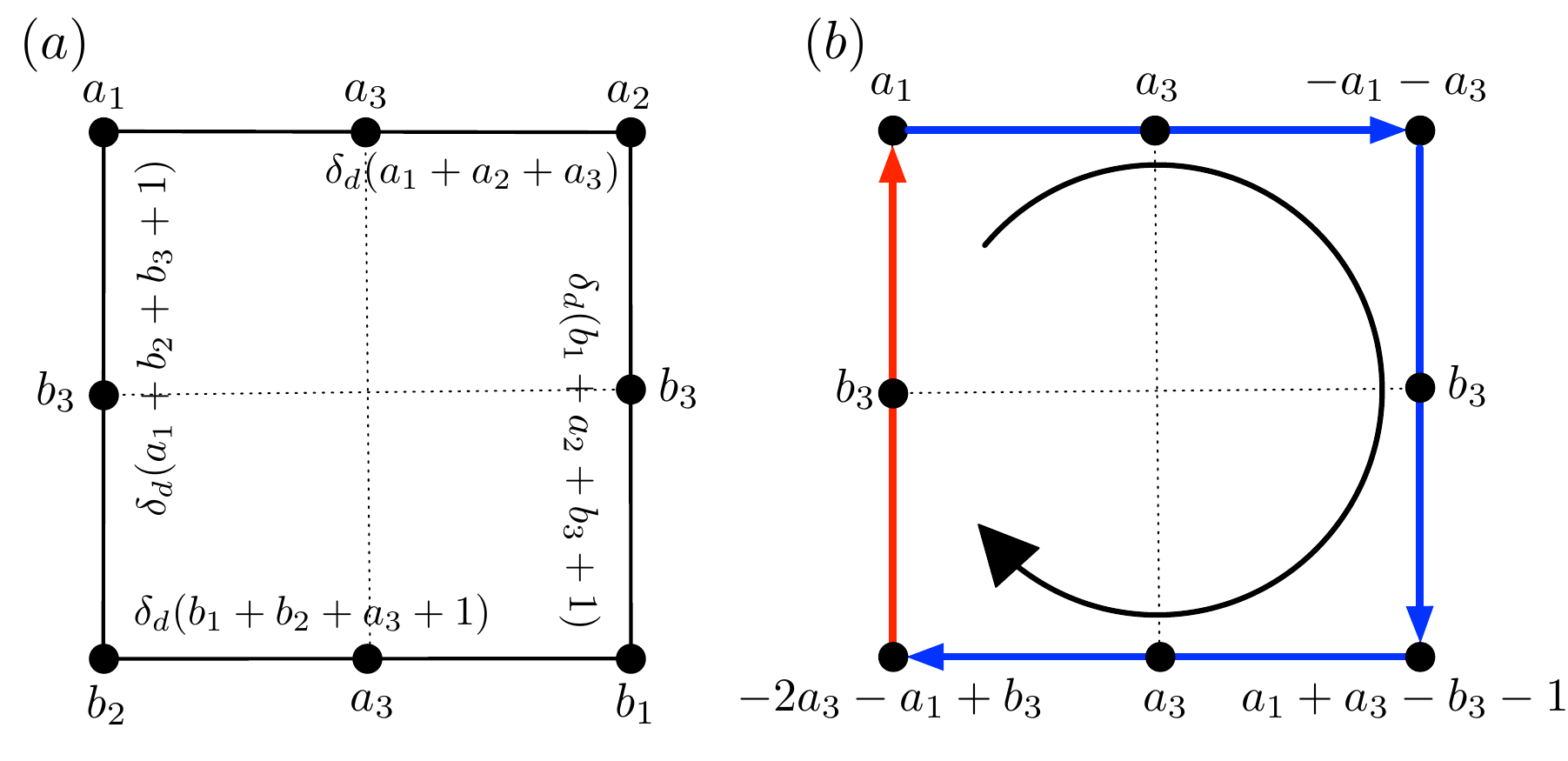}
  \caption{Geometrical representation of the generic Bell inequalities for two measurements settings. (a) Four delta functions $\delta_d (a_1 + a_2 + a_3)$, $\delta_d (b_1 + a_2 + b_3+1)$, $\delta_d (b_1 + b_2 + a_3+1)$, and $\delta_d (a_1 + b_2 + b_3+1)$ represented by solid lines connecting variables $a_i$ and $b_j$ at each vertex. (b) Value assignment showing that at most three delta functions can be simultaneously satisfied. The variables in the blue lines indicate solutions satisfying the corresponding delta functions from the left diagram, while following the circular path clockwise leads to the red line which represents a constraint that cannot be satisfied, proving that all four delta function cannot be satisfied simultaneously. This geometric approach demonstrates the classical upper bound as 3$d$/4 - 1.}
  \label{fig:2setting}
\end{figure}

Consider a square with eight vertices depicted in Fig.~\ref{fig:2setting}(a). Each vertex represents an integer value $a_j$ or $b_j$, and each solid line corresponds to one of the four delta functions in Eq.~(\ref{eq:LR_2}). Vertices connected by dotted lines are constrained to have identical values.
Integer assignments to satisfy the delta functions proceed in a clockwise sequence, as in Fig.~\ref{fig:2setting}(b). The systematic assignment demonstrates the maximum number of delta functions that can be simultaneously satisfied as follows:
\begin{enumerate}
\item The variables $a_1$ and $a_3$ serve as free parameters, while $a_2 = -a_1 -a_3$ ensures satisfaction of $\delta_d (a_1 +a_2 +a_3)=1$.
\item Subsequently, $b_1=a_1 +a_3 -b_3 -1$ is determined to satisfy $\delta_d (b_1 +a_2 +b_3 +1)=1$.
\item Parameter $b_2$ necessarily assumes the value $-2a_3 -a_1 +b_3$ to satisfy $\delta_d (b_1 +b_2 +a_3 +1)=1$. 
\item Finally, the delta function $\delta_d (a_1 + b_2 + b_3+1)$ becomes $\delta_d (-2a_3 + 2b_3 + 1)$, indicated by the red arrow.
\end{enumerate}
The last delta function cannot equal unity for even $d$-dimensional system by Theorem~\ref{theorem:ax=b}. The linear congruence $2(-a_3+b_3) +1 \equiv 0 \mod d$ has no solution for even $d$ because $g=\mathrm{gcd}(2, \mathrm{even}~d) = 2$ and $2 \nmid -1$, where the notation $a \nmid b$ indicates that $a$ does not divide $b$. Consequently, $\delta_d (-2a_3 + 2b_3 + 1) = 0$, thus the maximum number of delta functions that can simultaneously equal unity is $3$, establishing that $B_{2}^{\text{LHV}} \leq 3d/4 -1$.
\begin{theorem}
A linear congruence is a congruence relation of the form
\begin{equation}
aX \equiv b \mod m
\end{equation}
where $a,b,X,m \in \mathbb{Z}, a \neq 0, m > 0$. Let $g=\mathrm{gcd}(a,m)$. Then the linear congruence has a solution if  and only if $b$ is divisible by $g$.
\label{theorem:ax=b}
\end{theorem}
A key aspect of our method is the closed loop, wherein values are assigned in consecutive order to satisfy each delta function, ultimately yielding an incompatible constraint. This approach remains valid regardless of the initial assignment point, due to the permutation symmetries inherent in the linear congruence relations of the delta functions.

\begin{figure}[t]
\centering
\includegraphics[width=0.9\linewidth]{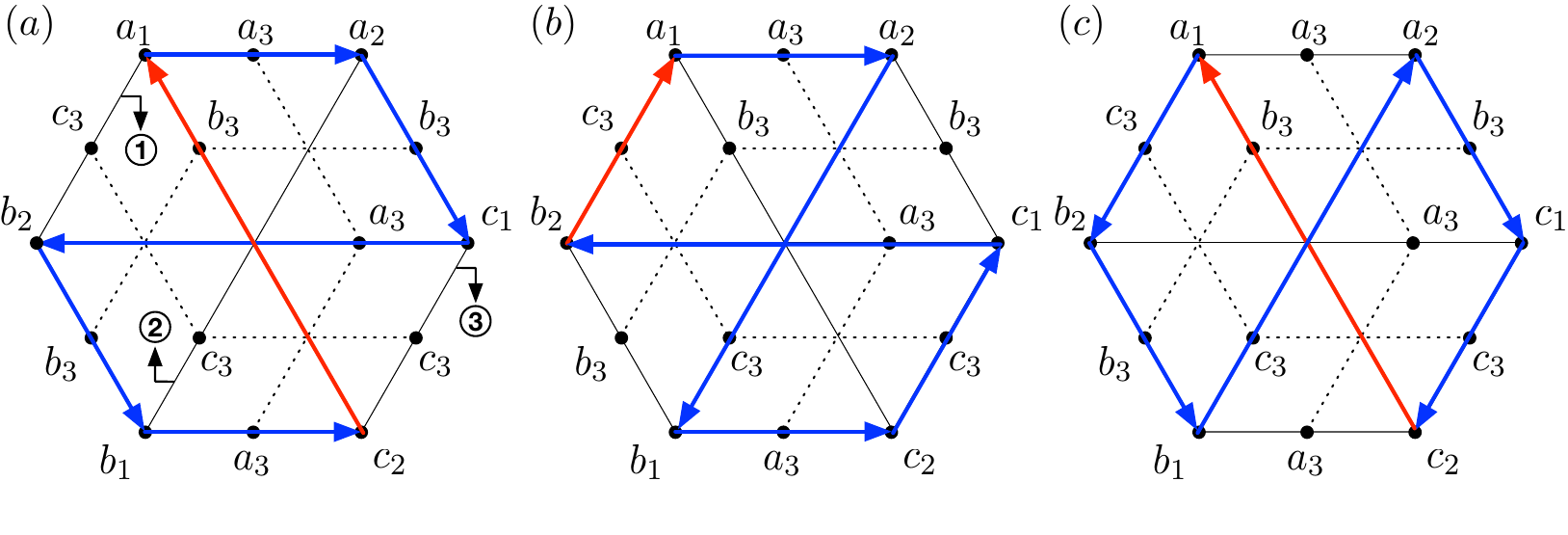}
\caption{Geometrical representation for calculating the classical upper bound with three measurement settings. Hexagonal representation of the nine delta functions, with vertices representing variables and dotted lines connecting identical values. (a)-(c) Three distinct loops generate the constraints, confirming that at most seven delta functions can simultaneously equal unity.}
\label{fig:3setting}
\end{figure}

Now we apply our geometric approach to the $(3,3,d)$ case. A hexagonal representation is employed to visualize the nine delta functions of $\mathcal{L}$ in Eq.~(\ref{eq:LR_3}), as depicted in Fig.~\ref{fig:3setting}. Similar to the two observables case, each vertex represents variables in the delta functions, with variables connected by dotted lines constrained to have identical values. The solid lines correspond to all delta functions of $\mathcal{L}$. Following the same methodology as in the two observables case, we identify three distinct closed loops, each composed of six delta functions with each loop imposing its own incompatible constraint. For the example of the first loop in Fig.~\ref{fig:3setting}(a), assignments for variables are as follows:
\begin{enumerate}
\item Set $a_2 = -a_1 -a_3$ to satisfy $\delta_d (a_1 +a_2 +a_3)=1$.
\item Set $c_1=a_1 -b_3 + a_3 -1$ to satisfy $\delta_d (c_1 +a_2 +b_3 +1)=1$.
\item Set $b_2=-a_1 + b_3 - 2a_3$ to satisfy $\delta_d (c_1 +b_2 +a_3 +1)=1$.
\item Set $b_1=a_1 -2b_3 + 2a_3 -1$ to satisfy $\delta_d (b_1 +b_2 +b_3 +1)=1$.
\item Set $c_2=-a_1 +2b_3 - 3a_3$ to satisfy $\delta_d (b_1 +c_2 +a_3 +1)=1$.
\item Finally, the delta function $\delta_d (a_1 + c_2 + b_3+1)$ is reduced to $\delta_d (-3a_3 + 3b_3 + 1)$, indicated by the red arrow.
\end{enumerate}
According to Theorem~\ref{theorem:ax=b}, the linear congruence of the last delta function,  $-3a_3 + 3b_3 +1 \equiv 0 \mod d$, has no solution for $d=3k$, where $k$ is an integer. This implies that $\delta_d(-3a_3 + 3b_3 +1)=0$, and consequently only five delta functions, indicated by the blue lines, can equal unity. This result indicates the quantum violation since it establishes $\mathcal{L} < 1$. However, to calculate the classical upper bound, we can examine the remaining three delta functions to obtain the exact upper bound. The remaining delta functions, denoted by circled numbers $\circled{1}$, $\circled{2}$, and $\circled{3}$, contain an undetermined variable $c_3$. With the assigned variables by the closed loop analysis, the three delta functions are reduced as follows: $\circled{1} \rightarrow \delta_d(-2a_3 + b_3 + c_3 + 1)$, $\circled{2} \rightarrow \delta_d(a_3 -2 b_3 + c_3)$, and $\circled{3} \rightarrow \delta_d(-2a_3 + b_3 + c_3 + 1)$. Note that no values of $a_3, b_3$, and $c_3$ exist for which all three delta functions equal to unity simultaneously, instead at most two delta functions can be satisfied concurrently. Consequently, there exists a solution of $a_j$, $b_j$, and $c_j$ that satisfies at most seven out of the nine delta functions. The classical upper bound of the generic Bell inequality for the $(3,3,d)$ system is $B_{3}^{\text{LHV}} \leq 7d/9 -1$, which contradicts the quantum upper bound $(d-1)$. The second and third loops in Fig.~\ref{fig:3setting}(b) and (c) yield the identical results, confirming that at most seven delta functions can simultaneously equal unity.

We now demonstrate the application of our geometric methodology to the generic Bell inequality involving four measurement settings, i.e., $(3,4,d)$ case. While the quantum violation for the $(3,4,d)$ system was previously examined in Ref.~\cite{ryu2008bell}, we present here an alternative derivation employing our geometric approach, which both confirms the previous finding and demonstrates the methodological advantages of geometric representation.
The generic Bell operator for $(3,4,d)$ system reads
\begin{equation}
\hat{B}_{4} = \frac{1}{4^3} \sum_{n=1}^{d-1} \sum_{\gamma=0}^{3} \bigotimes_{j=1}^
  {3} \sum_{\eta_{j}=0}^{3} \Omega^{\gamma \eta_{j}} \omega^{n \eta_{j} /4}
  \hat{X}_{j}^{n}(\eta_{j} /4),
\end{equation}
where $\Omega = \textrm{exp}(2 \pi i/4)$ and $\omega = \textrm{exp}(2 \pi i/d)$. It is remarkable that the tripartite generalized GHZ state $\ket{\psi}=\sum_{n=0}^{d-1} \ket{n,n,n}/\sqrt{d}$ is an eigenstate with eigenvalue $(d-1)$, i.e., $\bra{\psi}\hat{B}_{4}\ket{\psi} = d-1$. Note that the quantum upper bound for $(3,4,d)$ case is equivalent to the cases of $(3,2,d)$ and $(3,3,d)$. The Bell function based on LHVs for $(3,4,d)$ system can also be written in terms of the function $\mathcal{L}$ as
\begin{equation}
  B_{4}^{\text{LHV}}=\mathcal{L}d-1,
\end{equation}
where the function $\mathcal{L}$ is given by
\begin{equation}
  \mathcal{L} = \frac{1}{4^2} \sum_{\eta_{1},\eta_{2},\eta_{3}=0}^{3} \delta_4(\tilde{\eta}) \delta_d(\tilde{\eta} /4 + \tilde{\alpha}).
\end{equation}
\begin{figure}[t]
\centering
\includegraphics[width=\linewidth]{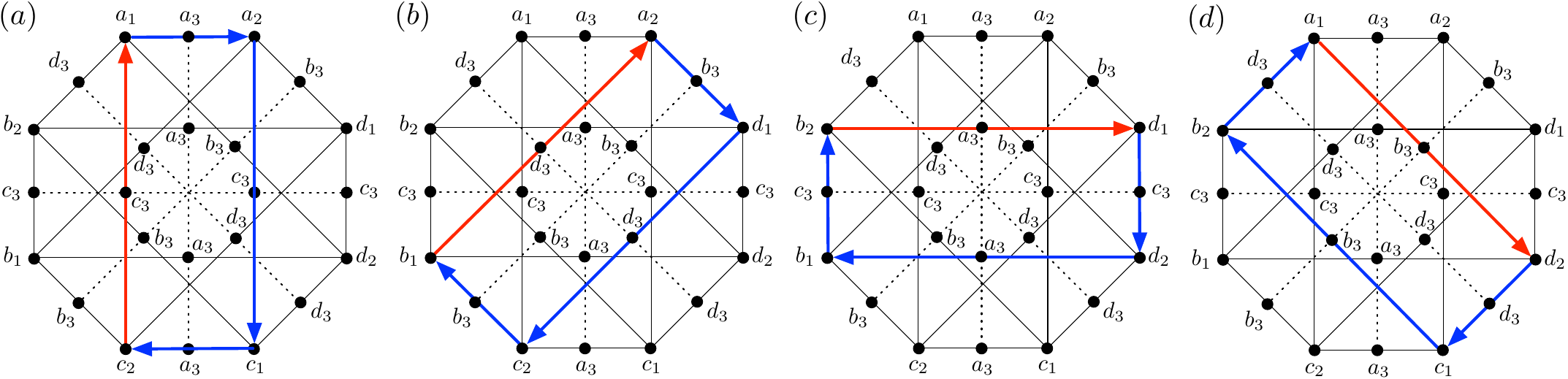}
\caption{Geometrical representation for calculating the classical upper bound with four measurement settings. All sixteen delta functions can be divided by four groups. As each group has the equivalent structure individually, the classical upper bound from one group is essential to calculate the total values of the classical upper bound.}
\label{fig:4set}
\end{figure}
From the function $\mathcal{L}$, the sixteen delta functions are given by
\begin{equation}
\begin{aligned}
\begin{cases}
  &\delta_d(a_1 + a_2 + a_3), \\
  &\delta_d(c_1 + a_2 + c_3+1) ,\\
  &\delta_d(c_1 + c_2 + a_3+1) ,\\
  &\delta_d(a_1 + c_2 + c_3+1),
\end{cases}
\end{aligned}\quad
\begin{aligned}
\begin{cases}
  &\delta_d(d_1 + a_2 + b_3+1) ,\\
  &\delta_d(d_1 + c_2 + d_3+2), \\
  &\delta_d(b_1 + c_2 + b_3+1) ,\\
  &\delta_d(b_1 + a_2 + d_3+1),
\end{cases}
\end{aligned}\quad
\begin{aligned}
\begin{cases}
  &\delta_d(d_1 + d_2 + c_3+2), \\
  &\delta_d(b_1 + d_2 + a_3+1) ,\\
  &\delta_d(b_1 + b_2 + c_3+1) ,\\
  &\delta_d(d_1 + b_2 + a_3+1), 
\end{cases}
\end{aligned}\quad
\begin{aligned}
\begin{cases}
  &\delta_d(c_1 + d_2 + d_3+2), \\
  &\delta_d(c_1 + b_2 + b_3+1) ,\\
  &\delta_d(a_1 + b_2 + d_3+1) ,\\
  &\delta_d(a_1 + d_2 + b_3+1). \nonumber
\end{cases}
\end{aligned}
\end{equation}
In order to apply the closed loop analysis, we employ an octagonal shape depicted in Fig.~\ref{fig:4set}. All sixteen delta functions are represented by the solid lines in Fig.~\ref{fig:4set}. Note that the sixteen delta functions can be categorized into four groups and each group has the same structure as that of $(3,2,d)$ system in Eq.~(\ref{eq:LR_2}). Therefore, the four delta functions in each group cannot all be unity for even $d$-dimensional system, with at most three delta functions being able to equal unity. Figures~\ref{fig:4set}(a)-(d) represent the closed loops corresponding to each group of the delta functions, and each red line indicates the constraint delta function. Consequently, the $\mathcal{L} =12/16=3/4$, and thus the classical upper bound of the generic Bell inequality for the $(3,4,d)$ system can be calculated as $B_{4}^{\text{LHV}} \leq 3d/4 -1$. This also can be violated by the quantum upper bound of $(d-1)$.

Finally, it would be interesting to consider how our geometric methodology can contribute to scenarios with arbitrary number of observables. The generic Bell operators for the $(3,M,d)$ case are given by
\begin{equation}
  \hat{B}_{M} = \frac{1}{M^3} \sum_{n=1}^{d-1} \sum_{\gamma=0}^{M-1} \bigotimes_{j=1}^{3} \sum_{\eta_{j}=0}^{M-1} \Omega^{\gamma \eta_{j}} \omega^{n \eta_{j} /M} \hat{X}_{j}^{n}(\eta_{j} /M),
  \label{eq:M_bell_operator}
\end{equation}
where $\Omega = \exp(2 \pi i/M)$ and $\omega = \exp(2 \pi i/d)$. Similar to our previous reasoning, the unitarity of the employed observables guarantees that the upper bound of the $\hat{B}_{M}$ is given by $(d-1)$. Moreover, the tripartite generalized GHZ state in Eq.~(\ref{eq:3_ghz}) achieves the upper bound, i.e., $\bra{\psi} \hat{B}_{M} \ket{\psi} = (d-1)$. This is because the mathematical formalism used to construct the generic Bell operators $\hat{B}_{M}$ ensures that they are designed to have the eigenvalue $(d-1)$ with respect to the $d$-dimensional GHZ state $\ket{\psi}$. To demonstrate a quantum violation, we need to prove that the LHVs cannot achieve the quantum maximum $(d-1)$, i.e., show $\mathcal{L}<1$, which reads
\begin{equation}
B_{M}^{\text{LHV}} = \frac{d}{M^{2}} \sum_{ \vec {\eta}=0}^{M-1} \delta_{M}(\tilde{\eta}) \delta_{d}(\tilde{\eta} /M + \tilde{\alpha}) -1 \equiv  \mathcal{L} d -1,
\end{equation}
with $\tilde{\eta} = \sum^{3}_{j=1}\eta_{j}$ and $\tilde{\alpha} = \sum^{3}_{j=1}\alpha(j,\eta_{j})$.
Finding a closed loop from our geometrical method can lead to the condition $\mathcal{L}<1$. The theoretical framework established in Ref.~\cite{ryu2014multisetting} proved the existence of the constraints in the form of linear congruences for arbitrary values of $M$. Specifically, these constraints exist when the dimension $d$ satisfies the relation $d=Mk$, where $k$ is an integer. This suggests that suitable geometric structures with closed loops can always be identified for arbitrary numbers of observables. Consequently, the quantum violations can be demonstrated in principle, although calculating precise classical upper bounds requires a more comprehensive analysis of the complete set of  delta function combinations and their corresponding constraint relations.

\section*{Conclusion}
We have demonstrated quantum violations of generic Bell inequalities involving multiple observables beyond the two-observable scenario. We constructed observable operators for each subsystem by employing quantum Fourier transformation and phase shift operations. Using these observables, we proposed generic Bell operators for systems with three and four observables per party, extending the original Son-Lee-Kim framework~\cite{son2006generic}. A key contribution of our work is the introduction of a geometric methodology that enables systematic calculation of the upper bounds based on local hidden variables for the generic Bell inequalities. Through this approach, we established quantum violations for the $(3,3,d)$ case when $d=3k$ and for the $(3,4,d)$ case with even dimensions $d$. Our approach can be extended to arbitrary values of $M$, where we expect the quantum violations to exist for dimensions satisfying $d=Mk$, with $k$ being an integer. Notably, our Bell operators consistently achieve a quantum upper bound of $(d-1)$ using the generalized Greenberger-Horne-Zeilinger state. Finally, we remark that it would be worth exploring the application of our geometric methodology to calculate the exact classical upper bounds for the generic Bell inequalities with arbitrary numbers of observables, which may further illuminate the fundamental boundary between quantum and classical physics.

\section*{Data availability}
The main data supporting the finding of this study are available within the article.

\bibliography{reference}

\section*{Acknowledgements}
This research was supported by Korea Institute of Science and Technology Information (KISTI).(No. K25L1M3C3) and by the National Research Foundation of Korea (NRF) (Grant No. RS-2023-NR119931 and RS-2023-NR119924).

\section*{Author contributions statement}
J.R. and J.L. developed the theoretical tools; All authors discussed the results and contributed to the writing of the manuscript.

\section*{Competing interests}
The authors declare no competing interests.

\section*{Appendix A: Derivation of the function $\mathcal{L}$}
We here derive the function $\mathcal{L}$ in Eq.~(\ref{eq:classical_function}). In local realistic
description, the generic Bell operator $\hat{B}_{3}$ in Eq.~(\ref{eq:3_bell_operator}) is replaced by the classical Bell function,
\begin{equation}
 B_{3}^{\text{LHV}} = \frac{1}{3^3} \sum_{n=1}^{d-1} \sum_{\gamma=0}^{2} \prod_{j=1}^{3} \sum_{\eta_{j}=0}^{2} \Omega^{\gamma \eta_{j}} \omega^{n \eta_{j} /3} \omega^{n \alpha(j,\eta_{j})},
\end{equation}
where by definition the outcome of measurement $X_j(\eta_j/3)$ is predetermined as its eigenvalue $\omega^{\alpha(j,\eta_j)}$, where $\alpha(j, \eta_{j})$ is integer. The classical Bell function $B_{3}^{\text{LHV}}$ is rewritten by noting the facts: $\sum_{\gamma=0}^{2} \Omega^{\gamma \eta_{j}}=3 \delta_3(\eta_j)$, where $\delta_3(\eta_j)=1$ if $\eta_j \equiv 0 \mod 3$ and $\delta_3(\eta_j) =0$ otherwise. We then obtain
\begin{eqnarray*}
  B_{3}^{\text{LHV}}
  &=& \frac{1}{3^{2}} \sum_{\eta_{1},\eta_{2},\eta_{3}=0}^{2} \delta_3(\tilde{\eta}) \sum_{n=1}^{d-1} \omega^{n(\tilde{\eta}/3 + \tilde{\alpha})},\\
  &=& \frac{1}{3^2} \sum_{\eta_{1},\eta_{2},\eta_{3}=0}^{2} \delta_3(\tilde{\eta}) \left[ d \delta_d(\tilde{\eta} /3 + \tilde{\alpha}) - 1 \right],\\
  &=& \frac{d}{3^2} \sum_{\eta_{1},\eta_{2},\eta_{3}=0}^{2} \delta_3(\tilde{\eta})\delta_d(\tilde{\eta} /3 + \tilde{\alpha}) - 1,
\end{eqnarray*}
where $\tilde{\eta} = \sum^{3}_{j=1}\eta_{j}$ and $\tilde{\alpha} = \sum^{3}_{j=1}\alpha(j,\eta_{j})$. Here we define the function $\mathcal{L}$ as 
\begin{equation}
  \mathcal{L} = \frac{1}{3^2} \sum_{\eta_{1},\eta_{2},\eta_{3}=0}^{2} \delta_3(\tilde{\eta})
  \delta_d(\tilde{\eta} /3 + \tilde{\alpha}).
\end{equation}

\section*{Appendix B: Eigenvalue equation for the generic Bell operator}
We show that for general $(N, M, d)$, the generic Bell operators have the generalized GHZ state as an eigenstate with an eigenvalue of $(d-1)$. Although our main text focuses on the case $N=3$, the derivation presented here holds for arbitrary $N$. The explicit form of the generic Bell operators in this case is given by
\begin{eqnarray*}
  \hat{B} = \frac{1}{M^N} \sum_{n=1}^{d-1} \sum_{\gamma=0}^{M-1} \bigotimes_{j=1}^{N} \sum_{\eta_{j}=0}^{M-1} \Omega^{\gamma \eta_{j}} \omega^{n \eta_{j} /M} \hat{X}_{j}^{n}(\eta_{j} /M),
\end{eqnarray*}
where $\Omega = \exp(2 \pi i/M)$, $\omega = \exp(2 \pi i/d)$, and the $n$-level raising operator $\hat{X}^n(\nu)$ shifts as
\begin{eqnarray*}
  \hat{X}^{ n}(\nu) \ket{m} = \begin{cases}
  \omega^{-\nu(n-d)} \ket{m+n} & \text{for $m \geq d-n$} \\
  \omega^{-\nu n} \ket{m+n} & \text{otherwise}.
  \end{cases}
\end{eqnarray*}
The $N$-partite and $d$-dimensional generalized GHZ state considered here is given by
\begin{eqnarray*}
\ket{\psi}=\frac{1}{\sqrt{d}} \sum_{k=0}^{d-1} \ket{k}^{\otimes N},
\end{eqnarray*}
where $\ket{k}^{\otimes N} \equiv \ket{k} \otimes \ket{k} \otimes \cdots \otimes \ket{k}$.
Using the definitions of the Bell operator and the GHZ state, the eigenvalue equation can be expanded as follows:
\begin{eqnarray*}
  \hat{B} \ket{\psi} &=& \frac{1}{\sqrt{d}}\frac{1}{M^N} \sum_{k=0}^{d-1}\sum_{n=1}^{d-1} \sum_{\gamma=0}^{M-1} \sum_{\vec{\eta}=0}^{M-1} \Omega^{\gamma \tilde{\eta}} \omega^{n \tilde{\eta}/M} \hat{X}_{1}^{n}(\eta_{1} /M) \ket{k} \otimes \hat{X}_{2}^{n}(\eta_{2} /M) \ket{k} \otimes \cdots \otimes \hat{X}_{N}^{n}(\eta_{N} /M)\ket{k},\\
&=& \frac{1}{\sqrt{d}}\frac{1}{M^{N-1}} \sum_{k=0}^{d-1}\sum_{n=1}^{d-1} \sum_{\vec{\eta}=0}^{M-1} \delta(\tilde{\eta} \equiv 0 \mod M) \omega^{n \tilde{\eta}/M} \hat{X}_{1}^{n}(\eta_{1} /M) \ket{k} \otimes \hat{X}_{2}^{n}(\eta_{2} /M) \ket{k} \otimes \cdots \otimes \hat{X}_{N}^{n}(\eta_{N} /M)\ket{k},\\
&=& \frac{1}{\sqrt{d}}\frac{1}{M^{N-1}} \sum_{\vec{\eta}=0}^{M-1} \delta(\tilde{\eta} \equiv 0 \mod M) \left[ \left(\omega^{\tilde{\eta}/M} \sum_{k=0}^{d-1}\hat{X}_{1}(\eta_{1} /M) \ket{k} \otimes \hat{X}_{2}(\eta_{2} /M) \ket{k} \otimes \cdots \otimes \hat{X}_{N}(\eta_{N} /M)\ket{k}\right)_{n=1}\right.\\
&+& \left(\omega^{2 \tilde{\eta}/M} \sum_{k=0}^{d-1}\hat{X}_{1}^{2}(\eta_{1} /M) \ket{k} \otimes \hat{X}_{2}^{2}(\eta_{2} /M) \ket{k} \otimes \cdots \otimes \hat{X}_{N}^{2}(\eta_{N} /M)\ket{k}\right)_{n=2} + \cdots\\
&+& \left.\left(\omega^{(d-1) \tilde{\eta}/M} \sum_{k=0}^{d-1}\hat{X}_{1}^{d-1}(\eta_{1} /M) \ket{k} \otimes \hat{X}_{2}^{d-1}(\eta_{2} /M) \ket{k} \otimes \cdots \otimes \hat{X}_{N}^{d-1}(\eta_{N} /M)\ket{k}\right)_{n=d-1}\right],
\end{eqnarray*}
where $ \sum_{ \vec {\eta}} \equiv \sum_{\eta_{1}} \cdots \sum_{\eta_{N}}$, $ \tilde{\eta} \equiv \sum_{j} \eta_{j}$, and we use $\sum_{\gamma=0}^{M-1} \Omega^{\gamma \tilde{\eta}} = M \delta(\tilde{\eta} \equiv 0 \mod M)$. Especially, the expression inside the square brackets $[\cdot]$ in the previous line can be further expanded as:
\begin{eqnarray*}
  && \omega^{\tilde{\eta}/M} \left(\omega^{-(1-d)\tilde{\eta}/M} \ket{0}^{\otimes N} + \omega^{-\tilde{\eta}/M}\sum_{k=0}^{d-2} \ket{k+1}^{\otimes N} \right) \\
  &+& \omega^{2\tilde{\eta}/M} \left(\omega^{-(2-d)\tilde{\eta}/M} \ket{0}^{\otimes N} + \omega^{-(2-d)\tilde{\eta}/M} \ket{1}^{\otimes N} + \omega^{-2\tilde{\eta}/M}\sum_{k=0}^{d-3} \ket{k+2}^{\otimes N}\right) + \cdots \\
  &+& \omega^{(d-1)\tilde{\eta}/M} \left(\omega^{\tilde{\eta}/M}\sum_{k=0}^{d-2} \ket{k}^{\otimes N} + \omega^{-(d-1)\tilde{\eta}/M} \ket{d-1}^{\otimes N}\right).
\end{eqnarray*}
Finally, we have the following form:
\begin{eqnarray*}
  \hat{B} \ket{\psi} &=& \frac{1}{\sqrt{d}}\frac{1}{M^{N-1}} \sum_{\vec{\eta}=0}^{M-1} \delta(\tilde{\eta} \equiv 0 \mod M) 
  \left[ \left(\omega^{d\tilde{\eta}/M} + \omega^{d\tilde{\eta}/M} + \cdots + \omega^{d\tilde{\eta}/M} \right) \ket{0}^{\otimes N}\right. \\
  &+& \left. \left(1 + \omega^{d\tilde{\eta}/M} + \cdots + \omega^{d\tilde{\eta}/M} \right) \ket{1}^{\otimes N} + \cdots + \left(1 + 1 + \cdots + 1\right) \ket{d-1}^{\otimes N}\right].
\end{eqnarray*}
Since $\sum_{\vec{\eta}=0}^{M-1}\delta(\tilde{\eta} \equiv 0 \mod M)$ restricts $\tilde{\eta}$ to be a multiple of $M$, there are $M^{N-1}$ such combinations. In this case, all $\omega^{x}$ terms reduce to $1$ due to $\omega = \exp(2\pi i /d)$, and the summation over the terms inside the parentheses yields a factor of $(d-1)$. Therefore, we obtain the following result:
\begin{eqnarray*}
\hat{B} \ket{\psi} = (d-1) \frac{1}{\sqrt{d}} \sum_{k=0}^{d-1} \ket{k}^{\otimes N} = (d-1) \ket{\psi}.
\end{eqnarray*}

\end{document}